# Electrochemically-driven insulator-metal transition in ionic-liquid-gated antiferromagnetic Mott-insulating NiS₂ single crystals

Sajna Hameed,[1,*,†] Bryan Voigt,[2] John Dewey,[2] William Moore,[2] Damjan Pelc,[1] Bhaskar Das,[2] Sami El-Khatib,[3,2] Javier Garcia-Barriocanal,[4] Bing Luo,[4] Nick Seaton,[4] Guichuan Yu,[4,5] Chris Leighton,[2,*] and Martin Greven[1,*]

[1]*School of Physics and Astronomy, University of Minnesota, Minneapolis, MN 55455, USA*
[2]*Department of Chemical Engineering and Materials Science,*
*University of Minnesota, Minneapolis, MN 55455, USA*
[3]*Department of Physics, American University of Sharjah, Sharjah, United Arab Emirates*
[4]*Characterization Facility, University of Minnesota, Minneapolis, MN 55455, USA*
[5]*Informatics Institute, University of Minnesota, Minneapolis, Minnesota 55455, USA*

**Abstract:** Motivated by the existence of superconductivity in pyrite-structure CuS₂, we explore the possibility of ionic-liquid-gating-induced superconductivity in the proximal antiferromagnetic Mott insulator NiS₂. A clear gating-induced transition from a two-dimensional insulating state to a three-dimensional metallic state is observed at positive gate bias on single crystal surfaces. No evidence for superconductivity is observed down to the lowest measured temperature of 0.45 K, however. Based on transport, energy-dispersive X-ray spectroscopy, X-ray photoelectron spectroscopy, atomic force microscopy, and other techniques, we deduce an *electrochemical* gating mechanism involving a substantial decrease in the S:Ni ratio (over hundreds of nm), which is both non-volatile and irreversible. This is in striking contrast to the reversible, volatile, surface-limited, *electrostatic* gate effect in pyrite FeS₂. We attribute this stark difference in electrochemical *vs.* electrostatic gating response in NiS₂ and FeS₂ to the much larger S diffusion coefficient in NiS₂, analogous to the different behaviors observed among electrolyte-gated oxides with differing O-vacancy diffusivities. The gating irreversibility, on the other hand, is associated with the lack of atmospheric S; this is in contrast to the better understood oxide case, where electrolysis of atmospheric $H_2O$ provides an O reservoir. This study of $NiS_2$ thus provides new insight into electrolyte gating mechanisms in functional materials, in a previously unexplored limit.

PHYSH:      Research areas:        Electrical properties

                 Physical systems:    Field-effect transistors, Mott insulators, antiferromagnets

                 Techniques:            Transport techniques, energy-dispersive x-ray

                                              spectroscopy, x-ray photoelectron spectroscopy, atomic

                                              force microscopy

*Corresponding authors: hamee007@umn.edu, leighton@umn.edu, greven@umn.edu

†Present address: Max Planck Institute for Solid State Research, Stuttgart 70569, Germany

# I.    Introduction

Chemical doping has long been a premier means to tune the charge-carrier density in insulators and semiconductors, thereby providing access to vast regions of the phase diagrams of these materials [1]. The use of electric field to manipulate the carrier density in transistor-type structures provides an attractive alternative to chemical doping. The advantages of this approach include the potential to electrostatically dope charge carriers with minimal associated chemical disorder, as well as continuous and reversible tuning of the charge carrier density. Conventional gate dielectrics such as $SiO_2$ enable the tuning of surface charge carrier densities in transistors only up to $10^{13}$ cm$^{-2}$ before breakdown, however [2]. Electric-double-layer transistors (EDLTs) that employ an ionic liquid (IL) or ion gel as the gate dielectric have emerged as an attractive alternative and have enabled charge-carrier-density tuning to well in excess of $10^{14}$ cm$^{-2}$ [3,4,5]. This has triggered several breakthroughs, including the discovery of superconductivity in $KTaO_3$ [6], gate-induced superconductivity in $SrTiO_3$ [7,8] and the high-$T_c$ cuprates [4,9,10], control of the insulator-metal transition in $VO_2$ [11,12] and $NdNiO_3$ [13,14,15], gating-induced ferromagnetism in diamagnetic $FeS_2$ [16], and electrostatic modulation of ferromagnetism in $La_{1-x}Sr_xCoO_{3-\delta}$ [17].

In early investigations with electrolyte-based transistor devices, the charge-carrier induction was thought to be purely electrostatic in nature, resulting in reversible control of electronic properties. However, several studies, particularly of oxides, have shown that electrolyte gating can also proceed *via* electrochemical mechanisms, through the formation/annihilation of anion vacancies [3,8,12,18,19,20,21,22], H$^+$ introduction [23,24,25], *etc*. It must be emphasized that such electrochemical mechanisms are not necessarily less favorable than electrostatic ones, and that they can in fact be advantageous due to broader property modulation [3]. Quite generally, in order to achieve *predictive* control of materials in such approaches, it is imperative to understand which properties of a material determine the extent of electrochemical *vs.* electrostatic response in electrolyte gating. While progress along these lines has been made in oxides [21,26], information from other materials systems is desirable.

The pyrite-structure first-row transition metal disulfides (TMS$_2$) exhibit a wide variety of magnetic and electronic properties [27], including diamagnetic semiconduction (FeS$_2$ [16,28], ZnS$_2$ [29]), antiferromagnetic Mott insulation (NiS$_2$ [30,31,32,33]), antiferromagnetism with a rare spin-state crossover (MnS$_2$ [34,35]), ferromagnetic metallicity (CoS$_2$ [36]), and superconductivity (CuS$_2$ [27,29]). Recent work even reported the discovery of Weyl fermions in CoS$_2$ [37]. It is also possible to tune these materials *via* substitutional chemical doping to obtain interesting properties such as half-metallic (or at least highly-spin-polarized) ferromagnetism in Co$_{1-x}$Fe$_x$S$_2$ [36,38,39] and metamagnetism in Co$_{1-x}$Ni$_x$S$_2$ [40]. Importantly, the structure of the series of TMS$_2$ compounds is cubic $Pa\bar{3}$, and the electronic and magnetic properties are controlled by $d$-band filling. These compounds are thus potentially ideal for exploration of possible gate-induced magnetism, insulator-metal transitions, and superconductivity.

A recent investigation of the influence of IL gating on the diamagnetic semiconductor FeS$_2$ (electronic configuration $t_{2g}^6 e_g^0$) revealed a transition to a ferromagnetic metallic state at positive gate bias [16], constituting the first demonstration of voltage-induced ferromagnetism from a diamagnetic state. Although semiconducting FeS$_2$ exhibits surface conduction that is extremely sensitive to surface structural and chemical modifications [41,42,43,44], the IL-gate-induced metallic state was observed to be strictly volatile and reversible (*i.e.*, the initial semiconducting state was recovered after bias removal), providing strong evidence that the gating mechanism in this material is a simple electrostatic one [16]. In NiS$_2$, surface conduction was recently shown to be prominent as well, with extreme sensitivity to surface modification with, *e.g.*, mechanical polishing [45]. $NiS_2$ adopts an antiferromagnetic Mott-insulating ground state with a Néel temperature of $T_N \approx 38$ K [31], and then undergoes a poorly understood first-order transition to a weak ferromagnetic state at $T_{wf} \approx 30$ K [31]. Prior work has shown that $NiS_2$ can be tuned to a metallic state with the application of pressure [46,47] or by substituting Se on the S site [47,48]. Superconductivity, however, has never been observed in doped $NiS_2$. In the phase diagram of the $TMS_2$ compounds, $NiS_2$ (electronic configuration $t_{2g}^6 e_g^2$) lies in close proximity to $CuS_2$ (electronic configuration $t_{2g}^6 e_g^3$), which exhibits superconductivity with a transition temperature of about 1.5 K [27,29]. $NiS_2$ is thus a fascinating candidate for possible gate-induced superconductivity, specifically at positive gate voltage, $i.e.$, in electron accumulation mode.

Here, we study the effect of IL gating on single-crystal $NiS_2$, exploring the possibility of gate-induced superconductivity. We find a clear gate-induced insulator-metal transition at positive gate bias, with progressively decreasing low-temperature sheet resistance with increasing gate voltage. No superconductivity is detected down to the lowest measured temperature of 450 mK, however. Most surprisingly, contrary to the highly reversible, volatile electrolyte-gate-induced surface insulator-metal transition in pyrite $FeS_2$ [16], transport, spectroscopy, and surface microscopy data on $NiS_2$ strongly implicate a non-volatile, irreversible electrochemical mechanism involving a substantial voltage-induced reduction in the S:Ni ratio. We find that the gate-induced metallic state is also three-dimensional (3D) in nature, with the decrease in S:Ni ratio occurring over depths of 100s of nm. We argue that this stark difference in gating mechanisms in $NiS_2$ and $FeS_2$ occurs due to the much larger S diffusion coefficient in $NiS_2$ [49] compared to $FeS_2$ [50,51]. This is analogous to the situation in electrolyte-gated oxides, wherein an electrostatic mechanism was established in materials such as $BaSnO_3$ due to low room temperature O vacancy diffusivity [26], in contrast to electrochemically-responding materials such as $La_{1-x}Sr_xCoO_{3-\delta}$ with high O diffusivity [20,21,22]. The irreversibility of the gating response in $NiS_2$ is then likely associated with the absence of an atmospheric S reservoir in IL gating of sulfides; this is fundamentally different from oxide IL gating, where electrolysis of atmospheric $H_2O$ (present in ILs) serves as an essentially limitless O reservoir.

## II.    Experimental Details

$NiS_2$ single crystals were grown *via* the chemical vapor transport method, as previously described [45]. Precursor powders of Ni (Alfa Aesar, 99.999% purity), S (CERAC, 99.9995% purity), and $NiBr_2$ (Sigma-Aldrich, 99.999% purity) were placed in sealed, evacuated (~$10^{-6}$ Torr), quartz tubes. Crystal growth then proceeded for 13 days in a two-zone tube furnace with hot and cold zones at 700 and 650 °C. The hot and cold zones were briefly inverted at the start of the growth to clean the growth zone. Post growth, crystals were washed in solvent to remove residual S and $NiBr_2$. Extensive structural, chemical, magnetic, and electrical characterization of these crystals was reported previously [45].

Four-terminal resistance measurements were carried out in a van der Pauw configuration in a Quantum Design PPMS Dynacool system in the temperature range 1.8 – 300 K. The samples were contacted by Al wire bonding onto gold pads sputtered onto the sample surface. A Keithley 2612B source-meter was used in a four-terminal configuration to source the measurement current and measure the voltage. A separate Keithley 2400 source-meter was used in a two-terminal configuration to apply a gate-voltage ($V_g$) with a concurrent measurement of the gate-current ($I_g$). The gate voltage was applied at 300 K for 30 minutes, before cooling the sample to the base temperature. The IL used for gating was EMI-TFSI [1-ethyl-3-methylimidazolium bis(trifluoro-methylsulfonyl) imide]. For low-temperature transport measurements to 0.45 K, we used a home-built $^3$He evaporation refrigerator with external gas handling. A high-throughput dipstick probe, directly inserted into the $^3$He pot, was used for these measurements. Unless otherwise stated, single crystals with a pristine (as-grown) top surface (on which the contacts were placed) and a mechanically polished bottom surface were used for all studies presented. This is important in light of the conclusions of our recent work on surface conduction [45], as discussed in more detail below.

Post-gating characterization was performed after removing the IL from the crystal by sequential sonication in acetone and ethanol for about 30 mins each. Chemical composition analysis was performed with a JEOL 6500 field-emission gun scanning electron microscope (FEG-SEM) equipped with energy-dispersive X-ray spectrometry (EDX). Incident electron energies in the 5 - 20 keV range were used. X-ray photoelectron spectroscopy (XPS) measurements were carried out in a PHI 5000 VersaProbe III photoelectron spectrometer (ULVAC-PHI) with a monochromatic Al K$\alpha$ X-ray source. The base pressure of the system was $4 \times 10^{-10}$ Torr, and the pressure during data collection was $7.5 \times 10^{-9}$ Torr. A spot diameter of 100 $\mu$m was utilized. The C 1$s$ peak was used as a binding energy reference, with its energy set to 284.8 eV. Low-energy Ar$^+$ ion-gun and electron-gun neutralizers were used to mitigate surface charging of the samples. For depth profiling, a 3 kV Ar$^+$ ion gun with a sputter area of $3 \times 3$ mm$^2$ and a sputter-rate of 4.3 nm/min (measured on SiO$_2$/Si) was used. Atomic force microscopy (AFM) was performed in contact mode on a Bruker Nanoscope V Multimode 8, and the data were analyzed using Gwyddion software [52,53].

### III.    Results and Analysis

A schematic of the EDLT geometry used to gate the $NiS_2$ single crystals is shown in the inset to Fig. 1(b). A gold-sputter-coated glass ring functions as both the gate electrode and as a container for the IL [16]. Single crystal $NiS_2$ samples with pristine (as-grown) top surfaces and polished bottom surfaces were used for the measurements, with the electrical contacts applied on the pristine top surfaces. As described in detail in our recent work [45], the surface conduction in these $NiS_2$ crystals is very sensitive to the surface preparation. For simplicity, we focus here only on gating pristine surfaces. Fig. 1(a) shows the temperature dependence of the sheet resistance ($R_s$) in a representative single-crystal $NiS_2$ EDLT at different applied $V_g$. At $V_g = 0$, typical semiconducting behavior is observed down to about 90 K, below which $R_s$ abruptly flattens, before increasing again at low $T$. This behavior arises from surface conduction in $NiS_2$ single crystals [32,45], where the more conductive surface shunts the insulating bulk at low $T$; this likely originates in surface states, which have been suggested to potentially be universal in $TMS_2$ compounds [45].

Upon application of only $V_g = +0.5$ V, a drastic decrease in the low-$T$ sheet resistance (two orders of magnitude at 30 K) is observed. In order to probe the volatility and reversibility of this gate-induced resistance change, this was followed by returning to $V_g = 0$ V and then applying -3.0 V. As seen in Fig. 1(a), the observed gate effect is completely non-volatile (compare +0.5 V with 0 V) and irreversible (compare to -3.0 V). These observations essentially rule out a simple electrostatic gating mechanism (as observed in $FeS_2$), immediately implicating electrochemistry, as returned to extensively below. Further application of progressively more positive $V_g$ then results in even stronger decreases in low-$T$ resistance, with $R_s$ eventually falling well below the 2D quantum resistance of ~26 k$\Omega$, to as low as ~10 $\Omega$ at $V_g = +2.0$ V. $dR_s/dT$ also becomes positive at this point, at least over some $T$ range. Further gating to higher $V_g$ (up to +4 V) does not significantly change $R_s$ (see Supplemental Material Section A [54]). We show below that these $R_s$ values far below 26 k$\Omega$ are in fact associated with strong reduction (*i.e.*, an electrochemical gating mechanism), occurring over length scales of 100s of nm into the crystal surface; the transition in Fig. 1(a) is thus from 2D insulator to 3D metal. An attempt to recover the ungated state by polishing the sample surfaces post-gating did indeed increase the low-$T$ resistance, but by barely an order-of-magnitude (see the dotted curve in Fig. 1(a)). This is due to the fact that the surfaces of polished $NiS_2$ are substantially more conductive than pristine surfaces [45], rendering this approach unproductive. We did confirm that the dramatic decrease in sheet resistance with increasing positive gate bias also occurs in samples with polished gated surfaces, however (see Supplemental Material Section B [54]). As a final comment on Fig. 1(a), note that these data were taken in a PPMS Dynacool system with a base temperature of 1.8 K. Additional transport experiments in a $^3$He refrigerator were performed on a crystal after gating to $V_g = +2.0$ V, but no superconductivity was detected down to 0.45 K (see the inset to Fig. 1(a)).

We also examined the temperature-derivative of the sheet resistance, as shown in Fig. 1(b). A clear anomaly is observed at $T \approx 30$ K and $V_g = 0$ V, which is well known to be associated with the weak ferromagnetic transition in $NiS_2$ at $T_{wf} \approx 30$ K [45,49]. This anomaly persists to $V_g = +0.9$ V, but then completely disappears upon further gating to $V_g = +1.3$ V and above. This observation is highly significant as the 30 K anomaly is known to be associated with the $NiS_2$ phase specifically; it does not occur in lower-S-content Ni sulfides, such as $NiS$, $Ni_3S_4$, *etc*. Additionally, the anomaly is observed to re-appear after polishing the sample surfaces post-gating (see dotted curve in Fig. 1 (b)). Taken together with the observed non-volatility and irreversibility of the gate effect, which strongly support an electrochemical (not electrostatic) mechanism, and the 3D nature of the gated metallic state, these observations clearly raise the possibility of reduction of the originally NiS$_2$ surface over significant depths, which we now explore with surface-sensitive chemical characterization techniques.

Fig. 2(a) shows EDX spectra obtained from the top surface of an NiS$_2$ single crystal, both before gating, and after gating to $V_g$ = +4.0 V and then removing the IL. Apart from the usual O and C signals due to surface contamination, the ungated sample shows only the expected Ni and S core transitions. In the gated sample, an additional weak F signal is visible, due to residual IL on the sample surface [20]. More importantly, the spectral intensities in Fig. 2(a) are normalized to those of the most intense Ni peak, the strong decrease in S intensity after gating therefore providing clear evidence of gating-induced reduction (*i.e.*, a decrease in the S:Ni ratio) in these data acquired at 5 keV incoming electron energy.

In EDX, the energy of the incident electron beam determines the depth to which the electrons penetrate, the generation volume within which the emitted photons originate, and thus the effective probe depth. This is illustrated in the inset to Fig. 2(b), which shows the depth below which 90% of the X-rays are ejected from the sample and detected, for varying incident electron energies (from the 5 to 20 keV studied in our experiments). These probing depths were calculated using the Monte Carlo simulation of electron trajectories available in the CASINO software package [55]. As shown in Fig. 2(b), varying the incident electron energy thus enables depth-profiling of the S:Ni ratio. At 20 keV, for example, where X-rays are ejected from depths up to ~1 $\mu$m (inset), the gated NiS$_2$ crystal has a S:Ni ratio of $1.92 \pm 0.10$ (systematic-error-dominated), corresponding to stoichiometric NiS$_2$ within error. As also shown in Fig. 2(b), this is essentially identical to the ungated crystal, consistent with our prior work [45]. With decreasing incident electron energy (and hence probing depth), however, a strong decrease in the S:Ni ratio is observed in the gated case, directly evidencing gate-induced reduction, with the effect being strongest at the surface. At 5 keV, for example (~200 nm probing depth (inset)), the measured S:Ni ratio of $1.19 \pm 0.10$ is drastically reduced from that of $NiS_2$. Note that the same sample surface measured at 5 keV before gating shows a S:Ni ratio of $1.96 \pm 0.10$, leaving no doubt that the low S:Ni ratio measured on the gated surface is induced by gating, rather than some systematic EDX error. A reduction of the S:Ni ratio to values between 1.1 and 1.3 at 5 keV incident electron energy was observed in all (four) gated samples gated to $V_g \geq +2.0$ V (data not shown). As already deduced indirectly from transport measurements, the gating effect in these $NiS_2$ crystals thus clearly results from an electrochemical mechanism associated with reduction at positive $V_g$, and not from electrostatic surface doping of electrons. This electrochemical reduction apparently takes place over surprisingly large depths, of order 100s of nm. This is unusual but not unprecedented in electrochemistry-based electrolyte gating. In perovskite cobaltites, for example, reduction at positive gate voltage can take place over more than 100 nm, enabled by a high O diffusion coefficient at room temperature [21,22].

Examination of the Ni-S phase diagram reveals that, with S reduction, the room-temperature-stable phases most likely to be induced with a S:Ni ratio between 2 and 1 are $Ni_3S_4$ (S:Ni = 1.33) and NiS (S:Ni = 1) [56]. Further decrease of the S:Ni ratio to values below 1 could potentially induce $Ni_9S_8$ and $Ni_3S_2$, although other metastable phases are also possible [56], as well as oxides and hydroxides, which we return to later. While our EDX results clearly establish strong reduction to S:Ni ratios of ~1.2 near the crystal surfaces after gating, this technique provides no information regarding the specific phases present. To understand this further we thus also carried out XPS measurements, which enable the determination of binding energies, and hence changes in the valence state of specific elements. Such a change in valence would be expected, *e.g.*, if conversion of $NiS_2$ to NiS occurs in the surface region during gating. In this specific case, both NiS$_2$ and NiS have Ni in the +2 valence state, with S valence states of (S$_2$)$^{2-}$ (*i.e.*, S$^{1-}$) and S$^{2-}$, respectively. This specific scenario would thus induce a change in S valence, but not in Ni valence. On the other hand, reduction to Ni$_3$S$_4$ would generate S$^{2-}$ with mixed 2+/3+ valence for Ni, a difference that ought to be distinguishable by XPS. Note here that the XPS probing depth is determined by the energy-dependent mean-free-path of the emitted photoelectrons, typically a few nm, rendering XPS far more surface sensitive than energy-dependent EDS.

Figs. 3(a) and (b) show XPS survey scans of NiS$_2$ crystal surfaces before gating and after gating to $V_g$ = +4.0 V, respectively. Various Ni and S spectral peaks are clearly discernible, in addition to expected contaminant peaks due to C and O. In Fig. 3(b), additional contaminant peaks from F and N are also visible, arising from residual IL [20]; stronger C and O contamination is also visible in (b), as might be expected. Figs. 3(c) and (d) then show high-resolution scans around the S 2$p$ and Ni 2$p_{3/2}$ peaks, respectively. The data in these figures were taken before gating (green), after gating to +4.0 V then removing the IL (blue), and additionally after one cycle (~1 min) of Ar$^+$-ion sputtering of the gated surface (red). The thickness of material removed by each cycle of Ar$^+$-ion sputtering was calibrated on a Si/SiO$_x$ substrate and found to be ~4 nm.

As shown in Fig. 3(c), the S 2$p_{3/2}$ and 2$p_{1/2}$ peaks in the ungated case occur at 162.2 eV and 163.5 eV, respectively, in good agreement with prior work on NiS$_2$ [57] and FeS$_2$ [58]. After gating (blue line), clear shifts of these S 2$p$ peaks to lower binding energy occur, along with a decrease in signal-to-noise ratio due to the residual IL and associated surface damage. Importantly, this is exactly as expected for conversion from S$^{1-}$ to S$^{2-}$, and is in fact very similar to observations for NiS and Fe$_{1-x}$S [57,58]. Note that an additional broad spectral component is observed in the post-gating S high-resolution scan at ~168 eV, very possibly arising from residual TFSI on the gated surface [59]. After Ar$^+$-ion sputtering for ~0.5 min (red line), the S 2$p$ peaks regain features somewhat reminiscent of the pristine $NiS_2$ surface, but retain distinct differences, specifically broader widths, slight binding energy shifts, and lesser splitting. It is not surprising that this level of sputtering does not return XPS spectra to the initial state given that electrochemical reduction occurs over 100s of nm, as deduced from EDS. The state after sputtering (red line) in fact appears intermediate between the initial (green) and post-gate/pre-sputter (blue) states. The S $2p$ peaks after gating also display both depth-wise and lateral variability (see Supplemental Material Section C), as returned to below.

Also significant, the primary Ni $2p_{3/2}$ peaks in Fig. 3(d) reveal no major change in binding energy after gating, despite the strong shifts seen for the S peaks in Fig. 3(c). This is as expected if the gating-induced phase were NiS, in which Ni remains in the same 2+ valence state as $NiS_2$. A higher binding energy hump does emerge after gating (blue line), however, and is decreased in intensity by sputtering (red line). Such a higher binding energy feature would be expected for $Ni_3S_4$, in which Ni has mixed 2+/3+ valence [60]. Combining the conclusions from S and Ni XPS, it is thus likely that both NiS and $Ni_3S_4$ form under electrochemical reduction, consistent with the post-gating surface S:Ni ratio from EDS of ~1.2, which is intermediate between NiS and $Ni_3S_4$. We wish to stress here that both depthwise and lateral inhomogeneity are likely inevitable in electrochemical gating, meaning that coexistence of NiS and $Ni_3S_4$ is entirely reasonable. Other phases such as oxides and hydroxides of Ni are also possible, which would also be consistent with presence of the higher binding energy feature in the Ni $2p_{3/2}$ spectrum [61,62]. As a final comment on Fig. 3, we note that depth-dependent studies of XPS spectra were also attempted, but were inconclusive with respect to the near-surface depth dependence of the S:Ni ratio, primarily due to strong effects of gating on the surface topography, as discussed below.

Before moving to surface topography, we first comment on the correspondence of the above EDS and XPS results with the transport data in Fig. 1. The most important point here is that both NiS and $Ni_3S_4$ can support metallic conductivity. NiS is known to exist in two forms: (*i*) hexagonal $\beta$-NiS exhibiting a transition to an antiferromagnetically-ordered state below $T_N \approx 264$ K, accompanied by a metal-semiconductor or metal-metal transition [63,64]; and (*ii*) rhombohedral $\gamma$-NiS exhibiting metallic conductivity [65]. Off-stoichiometric $\beta$-NiS and sintered $\gamma$-NiS exhibit metallic conductivity with low-$T$ resistivities that have been reported to reach as low as ~1 $\mu\Omega$cm [65,66]. $Ni_3S_4$ exhibits ferrimagnetism below $T_C \approx 20$ K and is metallic with low-$T$ resistivity at least as low as ~6.5 m$\Omega$cm [67]. Gated surfaces reaching resistances as low as ~10 $\Omega$ is therefore consistent with our EDS and XPS findings. As quantitative support for this, note that a sheet resistance of ~10 $\Omega$ due to a 100s-of-nm-thick layer corresponds to ~1 m$\Omega$cm resistivity, easily within reach of mixed phase NiS/$Ni_3S_4$ based on the above literature values.

Moving to surface topography, prior work on gated $La_{1-x}Sr_xCoO_{3-\delta}$ films, for example, revealed that in the high positive gate voltage regime where electrochemical mechanisms dominate, gating eventually induces etching, leading to the formation of pits on the film surface [20]. Such electrochemical etching has also been observed in FeSe films on $SrTiO_3$ and MgO substrates [68]. Given the strongly electrochemical nature of the gating uncovered here in $NiS_2$, it seems pertinent to investigate the effects of gating on the topography of the crystal surface. Figs. 4(a) and (b) thus show contact-mode AFM height images of the surface of an ungated $NiS_2$ crystal and a post-gating (+2.0 V) crystal, respectively. The surface topography is indeed distinctly different in the two cases. While approximately micron-scale lateral features are visible in both cases, these occur on different specific scales, at different frequencies, and are associated with different depths. Rich structure is in fact observed in Fig. 4(b) at short lateral length scales, as well as deep pits reminiscent of the etch pits seen in $La_{1-x}Sr_xCoO_{3-\delta}$.

Quantitative analysis of the surface height distribution was performed with the relevant power spectral density (PSD), defined as [69]:

$$W(k_x, k_y) = \frac{1}{L^2}\left[\sum_{m=1}^{N}\sum_{n=1}^{N} h_{mn} e^{-2\pi i \Delta L(k_x m + k_y n)}(\Delta L)^2\right]^2, \qquad (1)$$

where $L$ is the scan length along both the horizontal and vertical axes, $h_{mn}$ is the profile height at position $(m, n)$, $k_x$ and $k_y$ are the spatial frequencies along the $x$- and $y$-directions, $\Delta L$ is the distance between neighboring sampling points, and $N$ is the total number of sampling points along each direction (with $L = N\Delta L$). One-dimensional PSDs were then obtained by averaging along the direction perpendicular to the scan-direction (*i.e.*, along the $y$-axis in our case):

$$W_1(k) = \frac{1}{L}\sum_{k_y} W(k, k_y) \qquad (2).$$

Mathematically, the PSD thus represents the distribution of surface roughness with different associated spatial frequencies, as defined by the inverse wavelength of the topographic features. Thus, higher values of PSD at high frequencies indicate a higher density of low-wavelength features (and *vice versa*). The 1D-PSDs obtained for the ungated and gated crystal surfaces in Fig. 4(a,b) are shown in Fig. 4(c). Clear differences are apparent. Specifically, the gated PSD (blue) is distinctly shifted to higher spatial frequencies (lower wavelengths), with a knee forming at ~2.2 $\mu$m$^{-1}$ (vertical blue dashed line). This indicates a surface with dominant low-wavelength components, the extracted correlation length (from the shown fits, see Supplemental Material Section D [54]) falling by an order of magnitude with respect to the ungated case, from ~4.8 $\mu$m to ~450 nm. We believe that this finding is consistent with the above conclusions from spectroscopy. Specifically, conversion of the initial $NiS_2$ single crystal surface to a lower S:Ni phase such as NiS or $Ni_3S_4$ likely occurs with significant lateral inhomogeneity, with multiple phases present, potentially even with associated polycrystallinity. Under such circumstances, low-wavelength structure would be expected in AFM images, as is the case in Fig. 4(b).

Prior to discussing the origins and implications of our findings, we note that substantial additional efforts were devoted to using various forms of XRD to definitively identify the low S:Ni ratio majority phase(s) post-gating. This encompassed both lab-based measurements with an area detector and synchrotron-based reciprocal space mapping, as described in Supplemental Material Section E [54]. Likely hindered by the much larger penetration depth of the X-rays in such measurements compared to the thickness of the reduced layer, these attempts were not successful. It is also possible that the low S:Ni ratio phase(s) (and/or oxides and hydroxides) induced by gating are amorphous or weakly crystalline, which would obviously limit the usefulness of XRD.

## IV. Discussion

The above transport, chemical, and structural characterization measurements on electrolyte-gated $NiS_2$ single crystals clearly reveal an electrochemical gating mechanism, resulting in substantial decreases in the S:Ni ratio (from 2.0 to close to ~1.2) over 100s of nm depths. These gating-induced changes are both non-volatile and irreversible, in stark contrast with what is seen in isostructural pyrite $FeS_2$. In this section we provide a hypothesis for these differences, which we believe provides new insight into the general issue of understanding electrostatic *vs.* electrochemical response in electrolyte gated materials.

In the better understood case of electrolyte gating of binary and complex *oxides*, electrochemical response *via* oxygen vacancy formation/annihilation *vs.* electrostatic accumulation of electrons/holes has been demonstrated to depend strongly on the O vacancy ($V_O$) diffusion coefficient [21,26]. Specifically, recent electrolyte gating investigations of the perovskite oxide $BaSnO_3$, for example, revealed reversible and volatile control of transport properties across a remarkably wide gate-voltage window, which was argued to be a consequence of very low room temperature $V_O$ diffusivity [26]. In such circumstances, positive $V_g$ may induce $V_O$ formation at the extreme surface of the oxide, but the very slow diffusion prevents proliferation of $V_O$ to greater depths, thus minimizing electrochemical response. At the other extreme, oxides such as $La_{1-x}Sr_xCoO_{3-\delta}$, $SrTiO_3$, $VO_2$, *etc.*, have much higher $V_O$ diffusivity, enabling proliferation of $V_O$ to substantial depths, thereby promoting electrochemical response. In cobaltites, for example, the diffusion length for $V_O$ on the time and temperature scales relevant to typical electrolyte gating experiments can easily exceed 100 nm [21,22].

Extending the above arguments to pyrite-structure $TMS_2$ compounds is revealing. In particular, S and $V_S$ diffusion in $FeS_2$ is notoriously sluggish, the room-temperature-extrapolated S diffusion coefficient of $10^{-37}$ $m^2s^{-1}$ [50,51] yielding a diffusion length ($\sqrt{Dt}$ where $D$ is the diffusion coefficient and $t$ is time) of $\sim 10^{-17}$ m. In the above picture, this strikingly short length scale would promote an electrostatic electron doping mechanism, exactly as recently deduced from the volatile, highly reversible gate effect [16]. In $NiS_2$ on the other hand, for reasons that are not entirely clear but may be related to cell-volume expansion related to Mottness, the S diffusion coefficient has been estimated to be $\sim 10^{-10}$ $m^2s^{-1}$ at room temperature, *i.e.,* 27 orders of magnitude higher than in $FeS_2$ [49]. This radical difference yields a diffusion length for S vacancies in $NiS_2$ of $\sim 100$ $\mu$m, assuming 300 K and 30 mins. Critically, this is easily large enough to rationalize the 100s of nm length scales over which gate-induced electrochemistry is found to occur in the present work. Non-$NiS_2$ phases that arise as gating proceeds could well have lower S diffusion rates, but this simple estimate based on $NiS_2$ is nevertheless revealing. As a final note on this point we remark that while the method used to extract the diffusion coefficient in Ref. [49] is atypical, and should be treated only as an estimate, the enormous difference in $V_S$ diffusivity between $FeS_2$ and $NiS_2$ appears beyond question.

Finally, with an explanation for a strongly non-volatile, electrochemistry-based, and long-length-scale gating process in hand, we now turn to the observed irreversibility, *i.e.*, the inability to recover $NiS_2$ at negative $V_g$. While lower S and $V_S$ diffusivities in reduced phases such as NiS and $Ni_3S_4$ could play a role here, we believe that the primary issue is the lack of an atmospheric reservoir of S for re-sulfidation. In the better understood case of oxides it is widely believed that $H_2O$ present in ILs undergoes electrolysis at finite $V_g$ [3], providing an effectively limitless reservoir of O for reoxidation of reduced phases. In cobaltites for example, perovskite $SrCoO_3$ can thus be reduced to bownmillerite $SrCoO_{2.5}$ at positive bias, and then reoxidized to $SrCoO_3$ at negative voltage, in a reversible cycle [3,22,70,71,72]. In the case of sulfides, no such S reservoir is obviously available, which we believe plays the key role in strongly limiting reversibility, as is manifest in Fig. 1.

## V.      Summary and Conclusions

In conclusion, we have revealed a crossover from a 2D insulating phase to a 3D metallic phase in ionic-liquid-gated $NiS_2$ single crystals with increasing positive gate voltage. Despite the dramatic insulator-metal transition, no superconductivity was detected down to 450 mK. Of highest interest, the electrolyte gate effect is irreversible, non-volatile, and electrochemical in nature, proceeding *via* reduction of the S:Ni ratio to depths of 100s of nm. These conclusions are supported by electronic transport, spectroscopic, and surface microscopy studies, providing a detailed picture.

We explain these features in terms of unusually large sulfur diffusivity, which enables non-volatile reduction over large length scales, as well as the absence of an obvious sulfur reservoir in such gating, which severely limits reversibility. $NiS_2$ thus illuminates a heretofore unexplored limit of electrolyte gating.

## VI.     Acknowledgments

We thank Liam Thompson for help with electrical contact preparation. This work was supported primarily by the National Science Foundation through the University of Minnesota (UMN) MRSEC under Award number DMR-2011401. Parts of this work were carried out in the Characterization Facility, UMN, which receives partial support from the NSF through the MRSEC (Award Number DMR-2011401) and the NNCI (Award Number ECCS-2025124) programs.

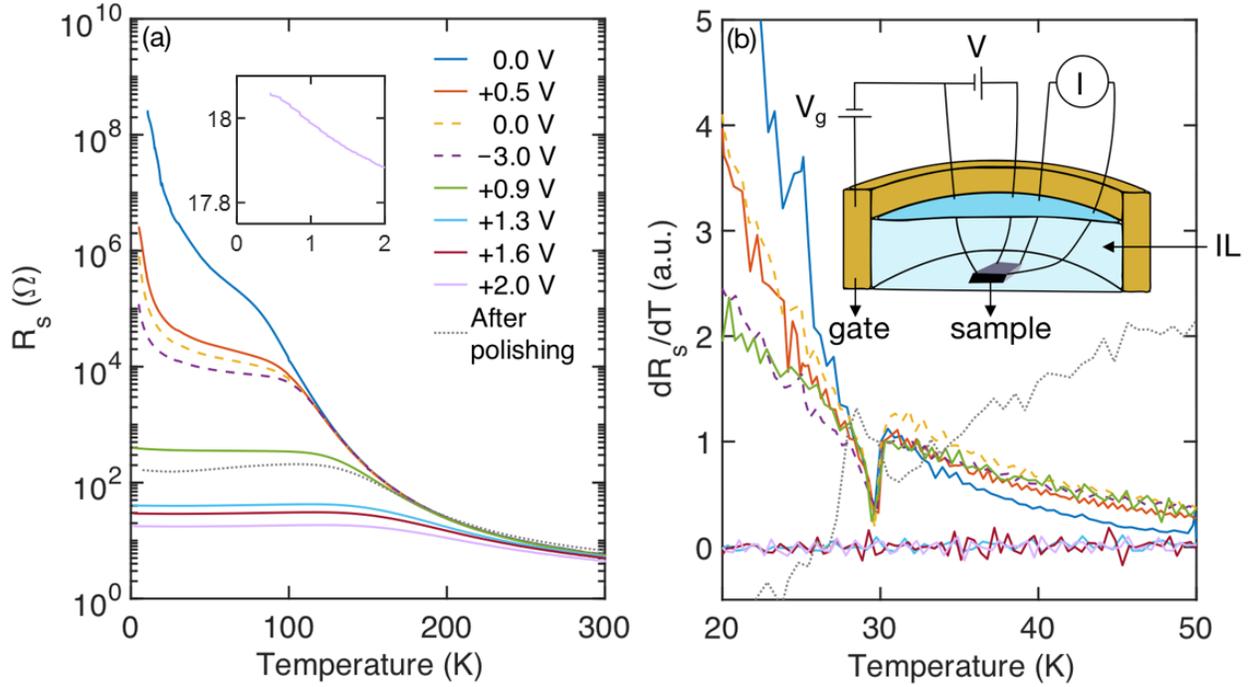

**Fig. 1.** (a) Sheet resistance as a function of temperature for different applied gate voltages. In the legend, the listing of applied voltages is in chronological order. As a check of the volatility and reversibility of the gate effect, 0 V and -3.0 V (dashed lines) were applied following the first positive gate voltage application of +0.5 V. The resistance of the sample after polishing all sides post-gating is also shown (dotted line). The inset shows low temperature $R_s$ vs $T$ obtained for the +2.0 V gated sample in a $^3$He refrigerator. (b) Temperature derivative of the sheet resistances in (a), highlighting the anomaly at about 30 K due to the onset of weak ferromagnetism. The data are normalized to the value at a temperature slightly above 30 K (in the cases where an anomaly occurs). The anomaly disappears at an applied gate voltage of +1.3 V, but reappears after the gated surfaces are removed by mechanical polishing. Inset: schematic of the gating setup.

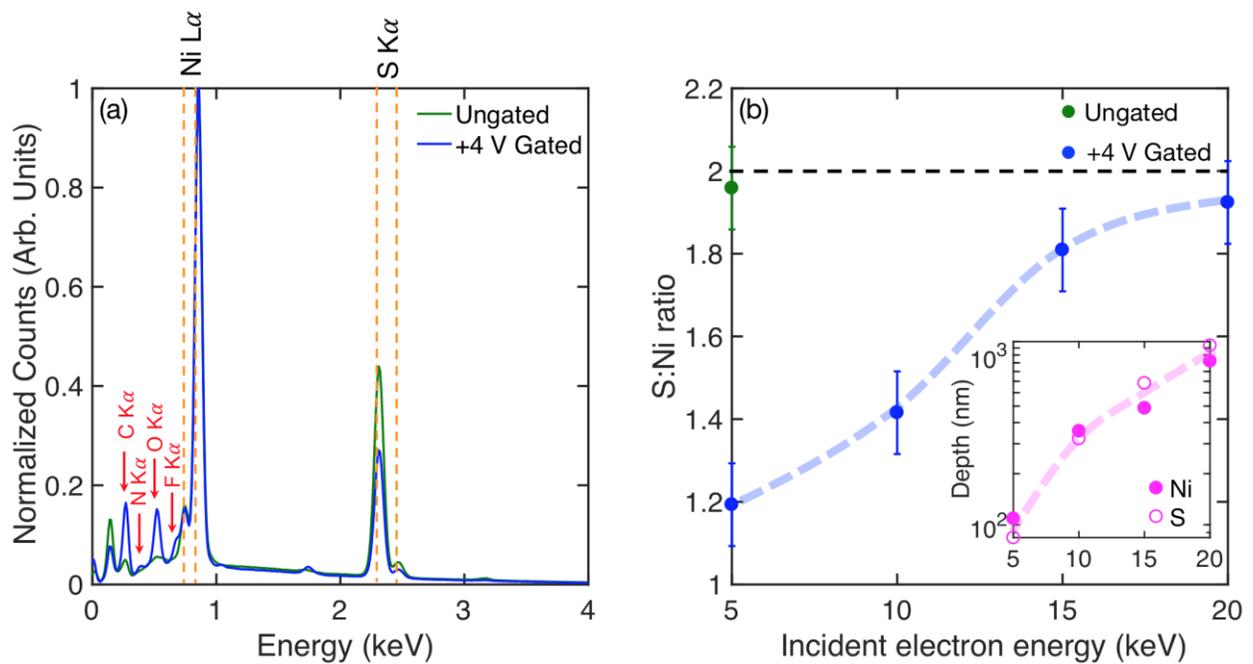

**Fig. 2.** (a) EDX analysis of a NiS$_2$ single crystal before and after gating (to +4.0 V). The spectra were obtained at an incident electron energy of 5 keV. The Ni and S peaks are marked by pairs of orange lines; common contaminant peaks are marked in red. The intensity is normalized to that of the most intense Ni peak. (b) S:Ni ratio obtained from EDX analysis of ungated and gated samples (+4.0 V) with different incident electron energies. The inset shows the depth below which 90% of the characteristic X-rays are emitted from the sample for S and Ni, as obtained from CASINO simulations (as discussed in the main text). The dashed lines in (b) are guides to the eye.

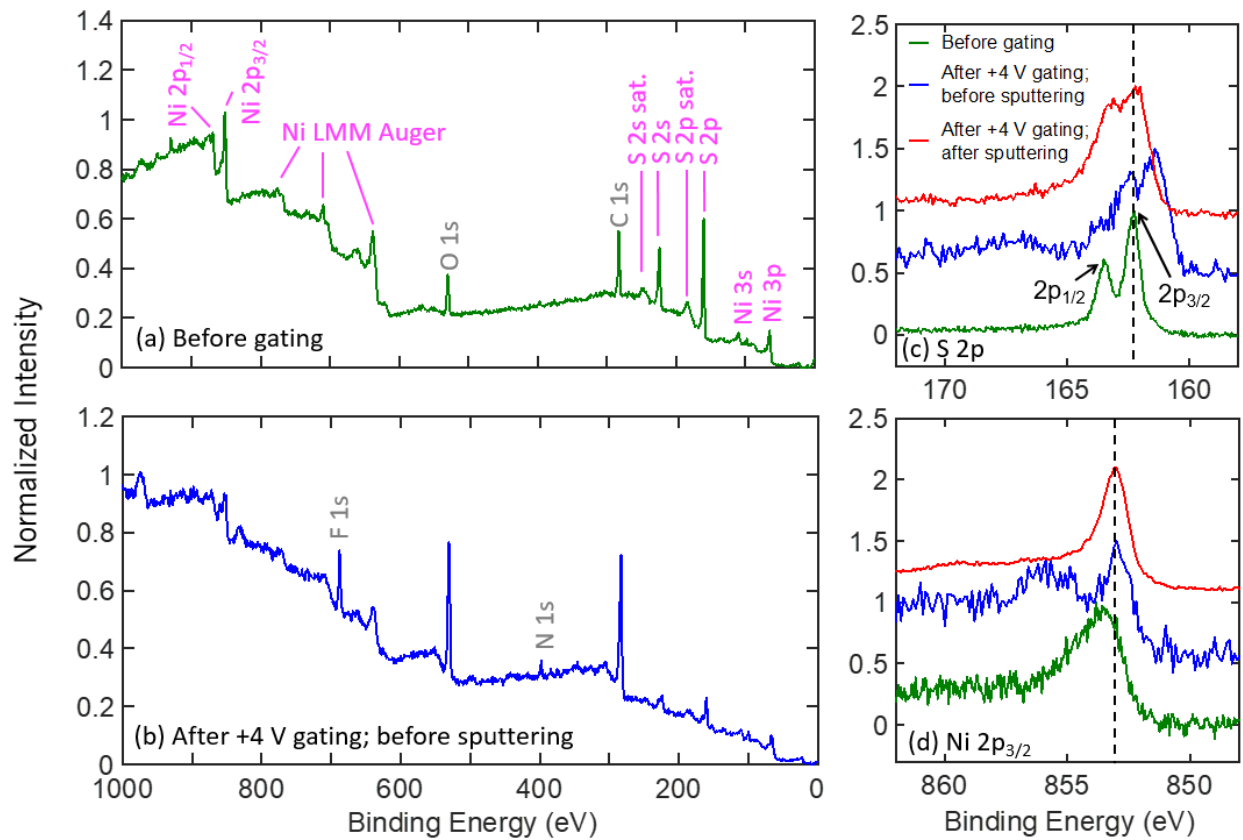

**Fig. 3.** XPS survey scans taken on the crystal surface (a) before gating (no IL applied) and (b) after gating (+4.0 V), but before Argon ion sputtering. The XPS peaks due to specific elements are marked in (a); in (b), only the peaks due to IL contamination are additionally marked. High-resolution XPS scans of the S 2*p* and Ni 2*p*$_{3/2}$ peaks are shown in (c) and (d), respectively. The spectra in green and blue correspond to panels (a) and (b), respectively. The spectra in red were obtained after sputtering ~2 nm off the surface with Argon-ion sputtering.

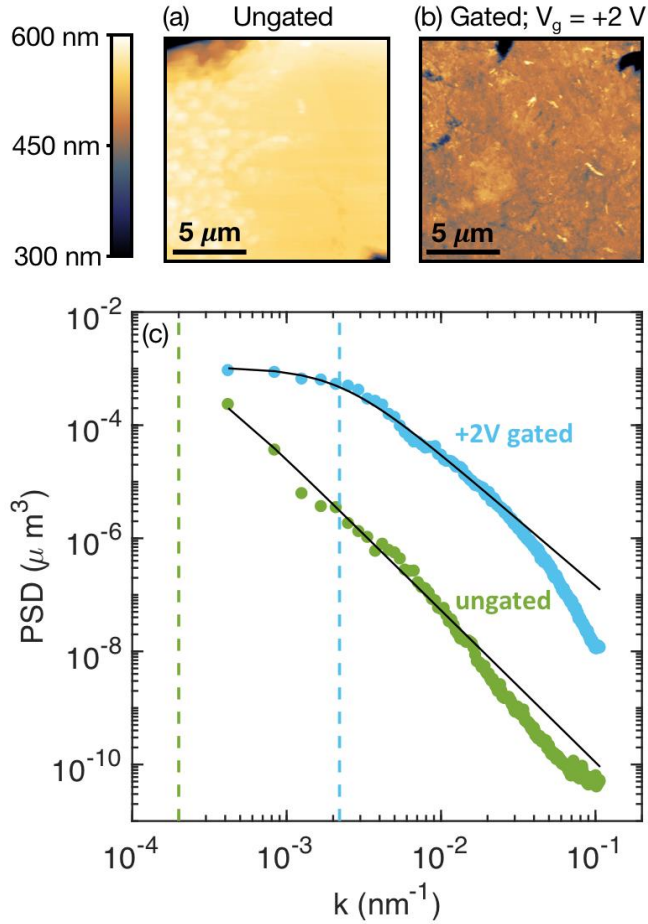

**Fig. 4.** Contact-mode AFM height images (15 $\mu$m × 15 $\mu$m) of (a) ungated and (b) +2.0-V-gated NiS$_2$ single crystal surfaces. The common height scale for the images is displayed on the left; the lower limit for the height scale was set at 300 nm, for better contrast. (c) 1D power-spectral density functions for gated (+2.0 V) and ungated samples, as obtained from the AFM height images in (a,b), using Eqs. (1) and (2). The solid black lines are fits to the $k$-correlation model (see Supplemental Material Section D [54]) used to extract the correlation length. The latter is 4.8(3.6) $\mu$m and 0.45(0.04) $\mu$m from the ungated and gated fits shown. The inverse correlation lengths obtained for the gated and ungated samples are depicted by dashed lines with the respective colors.

[1] M. Imada, A. Fujimori, and Y. Tokura, Rev. Mod. Phys. **70**, 1039 (1998).

[2] C. H. Ahn, J.-M. Triscone, and J. Mannhart, Nature **424**, 1015 (2003).

[3] C. Leighton, Nat. Mater. **18**, 13 (2019).

[4] A. M. Goldman, Annu. Rev. Mater. Res. **44**, 45 (2014).

[5] S. Z. Bisri, S. Shimizu, M. Nakano, and Y. Iwasa, Adv. Mat. **29**, 1607054 (2017).

[6] K. Ueno, S. Nakamura, H. Shimotani, H. T. Yuan, N. Kimura, T. Nojima, H. Aoki, Y. Iwasa, and M. Kawasaki, Nat. Nanotechnol. **6**, 408 (2011).

[7] K. Ueno, S. Nakamura, H. Shimotani, A. Ohtomo, N. Kimura, T. Nojima, H. Aoki, Y. Iwasa, and M. Kawasaki, Nat. Mater. **7**, 855 (2008).

[8] K. Ueno, H. Shimotani, Y. Iwasa, and M. Kawasaki, Appl. Phys. Lett. **96**, 252107 (2010).

[9] X. Leng, J. Garcia-Barriocanal, S. Bose, Y. Lee, and A. M. Goldman, Phys. Rev. Lett. **107**, 027001 (2011).

[10] A. T. Bollinger, G. Dubuis, J. Yoon, D. Pavuna, J. Misewich, and I. Božović, Nature **472**, 458 (2011).

[11] M. Nakano, K. Shibuya, D. Okuyama, T. Hatano, S. Ono, M. Kawasaki, Y. Iwasa, and Y. Tokura, Nature **487**, 459 (2012).

[12] J. Jeong, N. Aetukuri, T. Graf, T. D. Schladt, M. G. Samant, and S. S. P. Parkin, Science **339**, 1402 (2013).

[13] S. Asanuma, P. H. Xiang, H. Yamada, H. Sato, I. H. Inoue, H. Akoh, A. Sawa, K. Ueno, H. Shimotani, H. Yuan, M. Kawasaki, and Y. Iwasa, Appl. Phys. Lett. **97**, 142110 (2010).

[14] R. Scherwitzl, P. Zubko, I. G. Lezama, S. Ono, A. F. Morpurgo, G. Catalan, and J. M. Triscone, Adv. Mater. **22**, 5517 (2010).

[15] J. Son, B. Jalan, A. P. Kajdos, L. Balents, S. J. Allen, and S. Stemmer, Appl. Phys. Lett. **99**, 192107 (2011).

[16] J. Walter, B. Voigt, E. Day-Roberts, K. Heltemes, R. M. Fernandes, T. Birol, and C. Leighton, Sci. Adv. **6**, eabb7721 (2020).

[17] J. Walter, T. Charlton, H. Ambaye, M. R. Fitzsimmons, P. P. Orth, R. M. Fernandes, and C. Leighton, Phys. Rev. Mater. **2**, 111406(R) (2018).

[18] M. Li, W. Han, X. Jiang, J. Jeong, M. G. Samant, and S. S. P. Parkin, Nano Lett. **13**, 4675 (2013).

[19] T. D. Schladt, T. Graf, N. B. Aetukuri, M. Li, A. Fantini, X. Jiang, M. G. Samant, and S. S. P. Parkin, ACS Nano **7**, 8074 (2013).

[20] J. Walter, H. Wang, B. Luo, C. D. Frisbie, and C. Leighton, ACS Nano **10**, 7799 (2016).

[21] J. Walter, G. Yu, B. Yu, A. Grutter, B. Kirby, J. Borchers, Z. Zhang, H. Zhou, T. Birol, M. Greven, and C. Leighton, Phys. Rev. Mater. **1**, 071403(R) (2017).

[22] V. Chaturvedi, W. M. Postiglione, R. D. Chakraborty, B. Yu, W. Tabiś, S. Hameed, N. Biniskos, A. Jacobson, Z. Zhang, H. Zhou, M. Greven, V. E. Ferry, and C. Leighton, ACS Appl. Mater. Interfaces (2021).

[23] K. Shibuya and A. Sawa, Adv. Electron. Mater. **2**, 1500131 (2016).

[24] X. Leng, J. Pereiro, J. Strle, G. Dubuis, A. T. Bollinger, A. Gozar, J. Wu, N. Litombe, C. Panagopoulos, D. Pavuna, and I. Božović, Npj Quantum Mater. **2**, 35 (2017).

[25] H. Ji, J. Wei, and D. Natelson, Nano Lett. **12**, 2988 (2012).

[26] H. Wang, J. Walter, K. Ganguly, B. Yu, G. Yu, Z. Zhang, H. Zhou, H. Fu, M. Greven, and C. Leighton, Phys. Rev. Mater. **3**, 075001 (2019).

[27] S. Ogawa, J. Appl. Phys. **50**, 2308 (1979).

[28] A. Ennaoui, S. Fiechter, Ch. Pettenkofer, N. Alonso-Vante, K. Büker, M. Bronold, Ch. Höpfner, and H. Tributsch, Sol. Energy Mater. Sol. Cells **29**, 289 (1993).

[29] H. S. Jarrett, W. H. Cloud, R. J. Bouchard, S. R. Butler, C. G. Frederick, and J. L. Gillson, Phys. Rev. Lett. **21**, 617 (1968).

[30] K. Kikuchi, J. Phys. Soc. Japan **47**, 484 (1979).

[31] T. Thio, J. W. Bennett, and T. R. Thurston, Phys. Rev. B **52**, 3555 (1995).

[32] T. Thio and J. W. Bennett, Phys. Rev. B **50**, 10574 (1994).

[33] P. G. Niklowitz, P. L. Alireza, M. J. Steiner, G. G. Lonzarich, D. Braithwaite, G. Knebel, J. Flouquet, and J. A. Wilson, Phys. Rev. B **77**, 115135 (2008).

[34] J. M. Hastings and L. M. Corliss, Phys. Rev. B **14**, 1995 (1976).

[35] K. Persson, G. Ceder, and D. Morgan, Phys. Rev. B **73**, 115201 (2006).

[36] C. Leighton, M. Manno, A. Cady, J. W. Freeland, L. Wang, K. Umemoto, R. M. Wentzcovitch, T. Y. Chen, C. L. Chien, P. L. Kuhns, M. J. R. Hoch, A. P. Reyes, W. G. Moulton, E. D. Dahlberg, J. Checkelsky, and J. Eckert, J. Phys. Condens. Matter **19**, 315219 (2007).

[37] N. B. M. Schröter, I. Robredo, S. Klemenz, R. J. Kirby, J. A. Krieger, D. Pei, T. Yu, S. Stolz, T. Schmitt, P. Dudin, T. K. Kim, C. Cacho, A. Schnyder, A. Bergara, V. N. Strocov, F. de Juan, M. G. Vergniory, and L. M. Schoop, Sci. Adv. **6**, eabd5000 (2020).

[38] L. Wang, T. Y. Chen, C. L. Chien, J. G. Checkelsky, J. C. Eckert, E. D. Dahlberg, K. Umemoto, R. M. Wentzcovitch, and C. Leighton, Phys. Rev. B **73**, 1 (2006).

[39] L. Wang, K. Umemoto, R. M. Wentzcovitch, T. Y. Chen, C. L. Chien, J. G. Checkelsky, J. C. Eckert, E. D. Dahlberg, and C. Leighton, Phys. Rev. Lett. **94**, 056602 (2005).

[40] S. Ogawa, S. Waki, and T. Teranishi. Int. J. Magn. **5**:349 (2004).

[41] J. Walter, X. Zhang, B. Voigt, R. Hool, M. Manno, F. Mork, E. S. Aydil, and C. Leighton, Phys. Rev. Mater. **1**, 065403 (2017).

[42] M. Limpinsel, N. Farhi, N. Berry, J. Lindemuth, C. L. Perkins, Q. Lin, and M. Law, Energy Environ. Sci. **7**, 1974 (2014).

[43] M. Cabán-Acevedo, N. S. Kaiser, C. R. English, D. Liang, B. J. Thompson, H. Chen, K. J. Czech, J. C. Wright, R. J. Hamers, and S. Jin, J. Am. Chem. Soc. **136**, 17163 (2014).

[44] D. Liang, M. Cabán-Acevedo, N. S. Kaiser, and S. Jin, Nano Lett. **14**, 6754 (2014).

[45] S. El-Khatib, B. Voigt, B. Das, A. Stahl, W. Moore, M. Maiti, and C. Leighton, Phys. Rev. Mat. **5**, 115003 (2021).

[46] S. Friedemann, H. Chang, M. B. Gamza, P. Reiss, X. Chen, P. Alireza, W. A. Coniglio, D. Graf, S. Tozer, and F. M. Grosche. Sci. Rep. **6**:25335 (2016).

[47] J. M. Honig and J. Spalek, Chem. Mater. **10**, 2910 (1998).

[48] H. S. Jarrett, R. J. Bouchard, J. L. Gillson, G. A. Jones, S. M. Marcus, and J. F. Weiher, Mater. Res. Bull. **8**, 877 (1973).

[49] C. Clark and S. Friedemann, J. Magn. Magn. Mater. **400**, 56 (2016).

[50] E. B. Watson, D. J. Cherniak, and E. A. Frank, Geochim. Cosmochim. Acta **73**, 4792 (2009).

[51] Y. N. Zhang, M. Law, and R. Q. Wu, J. Phys. Chem. C **119**, 24859 (2015).

[52] http://gwyddion.net/

[53] D. Nečas, and P. Klapetek, Cent. Eur. J. Phys. **10**, 181 (2012).

[54] See supplemental material for additional gating data on $NiS_2$ single crystals with pristine top surfaces and polished bottom surfaces, gating data on double-side-polished $NiS_2$ single crystals, depth-profiles of a gated sample using XPS, quantitative analysis of PSD from AFM images, and X-ray diffraction data.

[55] https://www.gel.usherbrooke.ca/casino/

[56] H. Okamoto, J. Phase Equilibria Diffus. **30**, 123 (2009).

[57] P. Luo, H. Zhang, L. Liu, Y. Zhang, J. Deng, C. Xu, N. Hu, and Y. Wang, ACS Appl. Mater. Interfaces **9**, 2500 (2017).

[58] X. Zhang, T. Scott, T. Socha, D. Nielson, M. Manno, M. Johnson, Y. Yan, Y. Losovyj, P. Dowben, E. S. Aydil, and C. Leighton, ACS Appl. Mater. Interfaces **7**, 14130 (2015).

[59] C. Xu, B. Sun, T. Gustafsson, K. Edström, D. Brandell, and M. Hahlin, J. Mater. Chem. A **2**, 7256 (2014).

[60] J. Deng, Q. Gong, H. Ye, K. Feng, J. Zhou, C. Zha, J. Wu, J. Chen, J. Zhong, and Y. Li, ACS Nano **12**, 1829 (2018).

[61] A. P. Grosvenor, M. C. Biesinger, R. St. C. Smart, and N. S. McIntyre, Surface Science **600**, 1771 (2006).

[62] M .C. Biesinger, B. O. Payne, L. W. M. Lau, A. Gerson, and R. St. C. Smart, Surf. Interface Anal. **41**, 324 (2009).

[63] J. T. Sparks, and T. Komoto, Rev. Mod. Phys. **40**, 752 (1968).

[64] P. Chen, and Y. W. Du, Europhys. Lett. **53**, 360 (2001).

[65] L. J. Pauwels, and G. Maervoet, Bull. Soc. Chim. Belges, **80**, 501 (1971).

[66] L. J. Pauwels, and G. Maervoet, Bull. Soc. Chim. Belges, **81**, 385 (1972).

[67] A. Manthiram, and Y. U. Jeong, J. Solid State Chem. **147**, 679 (1999).

[68] J. Shiogai, Y. Ito, T. Mitsuhashi, T. Nojima, and A. Tsukazaki, Nat. Phys. **12**, 42 (2016).

[69] S. J. Fang, S. Haplepete, W. Chen, C. R. Helms, and H. Edwards, J. Appl. Phys. **82**, 5891 (1997).

[70] N. Lu, P. Zhang, Q. Zhang, R. Qiao, Q. He, H.-B. Li, H.-B. Li, Y. Wang, J. Guo, D. Zhang, Z. Duan, Z. Li, M.Wang, S. Yang, M. Yan, E. Arenholz, S. Zhou, W.Yang, L. Gu, C.-W. Nan, J.Wu, Y. Tokura, and P. Yu, Nature **546**, 124 (2017).

[71] Q. Lu, S. Huberman, H. Zhang, Q. Song, J. Wang, G. Vardar, A. Hunt, I. Waluyo, G. Chen, and B. Yildiz, Nat. Mater. **19**, 655 (2020).

[72] S. Ning, Q. Zhang, C. Occhialini, R. Comin, X. Zhong, and C. A. Ross, ACS Nano **14**, 8949 (2020).

Supplemental Material for:

# Electrochemically-driven insulator-metal transition in ionic-liquid-gated antiferromagnetic Mott-insulating NiS₂ single crystals

Sajna Hameed,[1,*,†] Bryan Voigt,[2] John Dewey,[2] William Moore,[2] Damjan Pelc,[1] Bhaskar Das,[2] Sami El-Khatib,[3,2] Javier Garcia-Barriocanal,[4] Bing Luo,[4] Nick Seaton,[4] Guichuan Yu,[4,5] Chris Leighton,[2,*] Martin Greven[1,*]

[1]*School of Physics and Astronomy, University of Minnesota, Minneapolis, MN 55455, USA*
[2]*Department of Chemical Engineering and Materials Science,*
*University of Minnesota, Minneapolis, MN 55455, USA*
[3]*Department of Physics, American University of Sharjah, Sharjah, United Arab Emirates*
[4]*Characterization Facility, University of Minnesota, Minneapolis, MN 55455, USA*

**Section A: Additional Gating Data on NiS$_2$ Single Crystal Samples with Pristine Top Surfaces and Polished Bottom Surfaces**

In this section, we document $R_s$ *vs.* $T$ curves for a few different gated samples (see Fig. S1). As pointed out in the main text, gating beyond $V_g = +2.0$ V does not significantly change $R_s$, compared to samples gated up to $V_g = +2.0$ V. The low temperature sheet resistances are observed to fall in the 10 - 100 $\Omega$ range for all samples gated to $V_g \geq +2.0$ V.

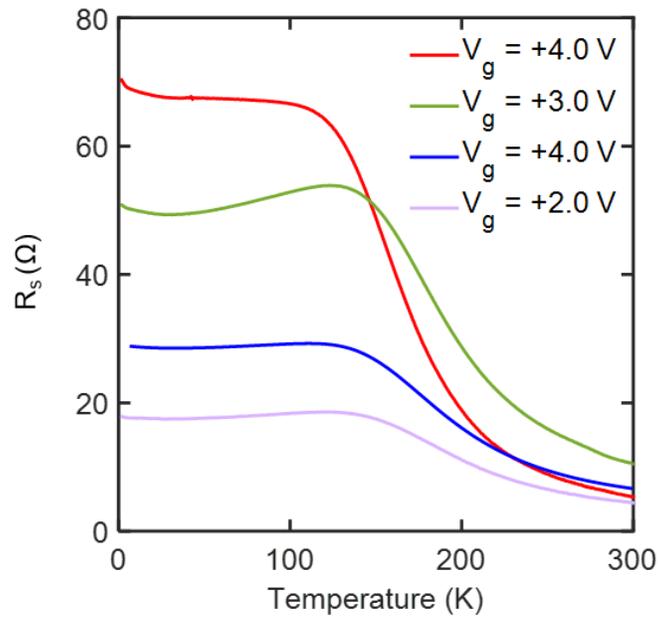

**Fig. S1**: Sheet resistance as a function of temperature for four different samples gated to the indicated gate voltages. The low temperature sheet resistances of all samples are observed to fall in the 10 - 100 $\Omega$ range. The $V_g = +2.0$ V gated crystal is the same as the one in Fig. 1 of the main text.

**Section B: Gating Data on Double-Side-Polished NiS₂ Single Crystals**

As pointed out in the main text, gating was also performed on samples with mechanically polished top and bottom surfaces, *i.e.*, double-side-polished crystals as opposed to the single-side-polished (pristine top surface/polished bottom surface) crystals in the main text. Fig. S2 shows an example of gating data on a double-side-polished crystal. As discussed in the main text, the initial state is more conductive when the top surface is polished, but the dramatic decrease in low temperature sheet resistance is still observed with increasing positive gate bias, similar to samples with pristine surfaces (Fig. 1).

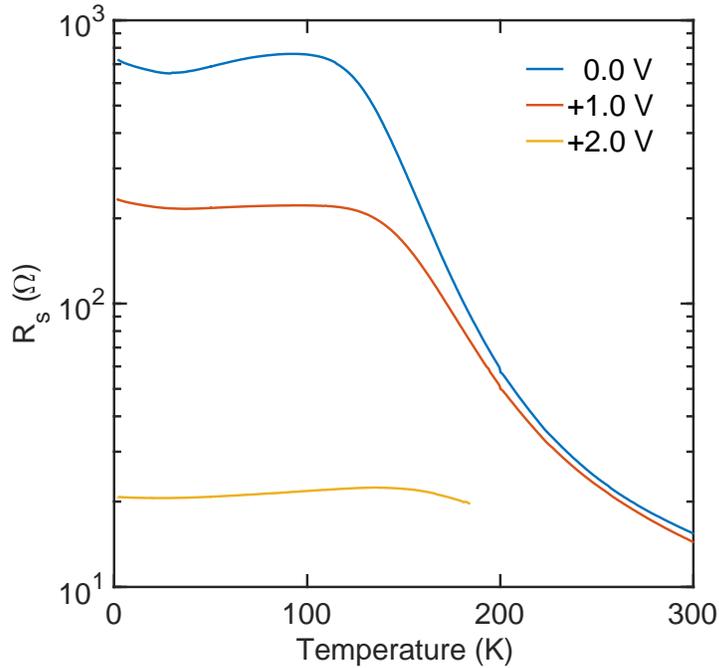

**Fig. S2**: Sheet resistance as a function of temperature at different applied gate voltages in a sample with mechanically polished top and bottom surfaces. The legend is in chronological order of applied gate voltages. A strong reduction in sheet resistance is observed with increasing positive gate bias, similar to that observed in pristine samples (Fig. 1(a)). The temperature range at $V_g$ = +2.0 V is cut off at ~190 K due to a contact failure.

**Section C: Depth Profile of a Gated Sample Using X-ray Photoelectron Spectroscopy**

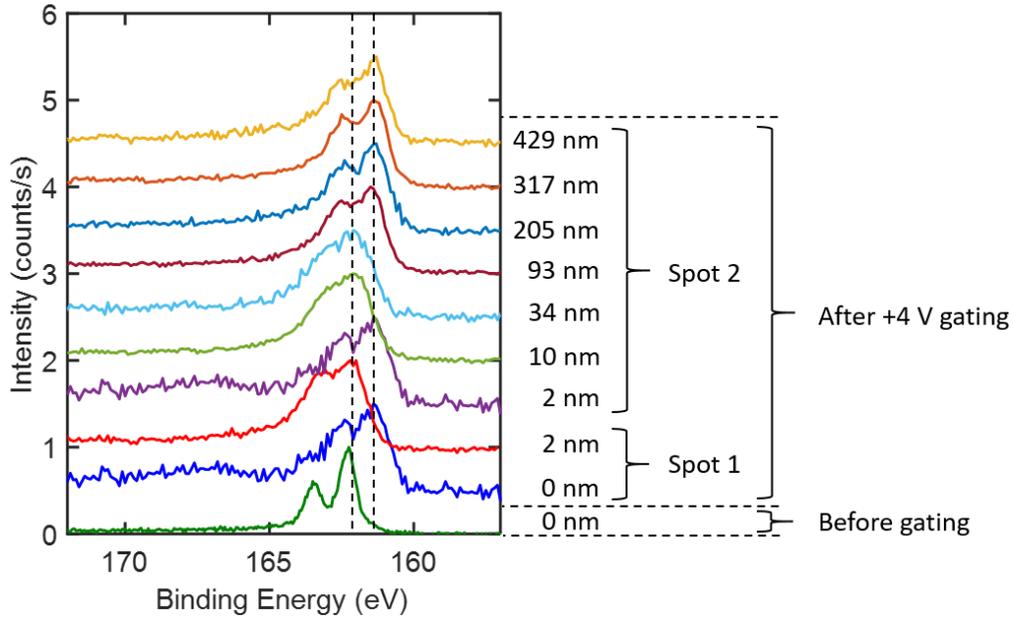

**Fig. S3**: High-resolution XPS scans of S 2*p* peaks obtained for a sample before gating and after gating to +4.0 V (the same sample as in Fig. 3 of the main text). Post gating, two different spots are measured. Argon-ion sputtering is used to depth-profile the gated sample at these two different spots. The thickness of the material sputtered before measurement of each spectra is indicated. Note that the indicated sputtered thicknesses are based on sputter rates that were calibrated on a Si/SiO$_x$ substrate and could be somewhat different for the NiS$_2$ surface.

In this section, we discuss additional XPS spectra obtained from the same sample as in Fig. 3 of the main text. Two different spots on the sample surface were measured after gating. The results from measurements at the first spot (Spot 1) were already presented in Fig. 3 of the main text and are reproduced in Fig. S3. We also measured a second spot (Spot 2) where we carried out more extensive depth profiling (Fig. S3). As discussed in the main text, the S 2*p* spectrum obtained at Spot 1 after gating is observed to be shifted to lower binding energies compared to the S 2*p* spectrum before gating, indicating a reduction of the S valence state. Upon sputtering the surface by ~ 2 nm, the S 2$p$ spectrum at Spot 1 shifts back to the initial position, but remains considerably broadened compared to the S 2$p$ spectrum before gating. On the other hand, measurements at Spot 2 reveal a S 2$p$ spectrum that is distinctly different from that measured at Spot 1. In fact, it appears similar to the shifted (reduced S) spectrum observed at Spot 1 before sputtering. This indicates that the S reduction induced by gating is laterally inhomogeneous. Moreover, further sputtering and measuring at Spot 2 reveal spectra that shift back and forth between spectra that resemble the ungated surface and spectra that resemble the reduced S state. This indicates that the S reduction induced by gating is depth-wise inhomogeneous as well.

**Section D: Quantitative Analysis of Power Spectral Density (PSD) From Atomic Force Microscopy (AFM) Images**

As explained in the main text, a quantitative analysis of the PSD data in Fig. 4(c) was performed to extract correlation lengths. A good fit to the frequency dependence of the PSD in Fig. 4(c) was obtained using a *k*-correlation model (also known as the ABC model) [1], defined by:

$$W_1(K) = \frac{A}{(1 + B^2 K^2)^{C/2}} \tag{3}.$$

Here, the parameter *A* is the low spatial-frequency limit of the spectrum, *B* determines the position of the 'knee' and defines a height-height correlation length beyond which the surface height fluctuations are uncorrelated, and *C* is related to the average fractal dimension. The knee in the gated crystal data therefore indicates a surface with dominant low-wavelength components, as would be expected if a gating-induced polycrystalline reduced S content phase formed on the single-crystal surface.

| Sample | $A$ ($\mu$m$^3$) | $B$ ($\mu$m) | $C$ |
|---|---|---|---|
| ungated | $1.8(3.2) \times 10^{-3}$ | 4.8(3.6) | 2.69(8) |
| gated | $1.1(0.1) \times 10^{-4}$ | 0.45(0.04) | 2.33(9) |

**Table S1**: Fit parameters from the *k*-correlation model, obtained for the PSD data in Fig. 4(c).

Table S1 summarizes the fit parameters obtained for the gated and ungated samples. Large errors are observed for the fit parameters corresponding to the ungated sample, primarily because the absence of the 'knee' feature leads to a large value for the parameter *B*. While the surface roughness, which is proportional to $\sqrt{A/B}$, and the fractal dimension *C* of the gated and ungated samples are comparable, the correlation length $B$ is observed to be an order of magnitude smaller in the gated sample, consistent with the interpretation of formation of a possibly polycrystalline reduced S content phase on the surface.

**Section E: X-ray Diffraction Attempts to Identify Gating-Induced Reduced S Content Phases**

As alluded to in the main text, X-ray diffraction (XRD) measurements were performed to attempt to further identify the reduced S content phases induced by gating. These measurements consisted of both synchrotron-based measurements using beamline 6ID-D of the Advanced Photon Source, Argonne National Laboratory and lab-based measurements on a Bruker D8 Discover X-ray diffractometer equipped with a Co K$\alpha$ X-ray source and a VANTE-500 area detector. As explained below and alluded to in the main text, neither approach was successful. Due to the reflection geometry of the lab-based approach, compared to transmission for the synchrotron measurements, we consider the former the most useful and present here only the results from those measurements. The measurements were performed in reflection with incident angles in the range 12.5° to 35° for the ungated sample and 12.5° to 37.5° for the gated sample.

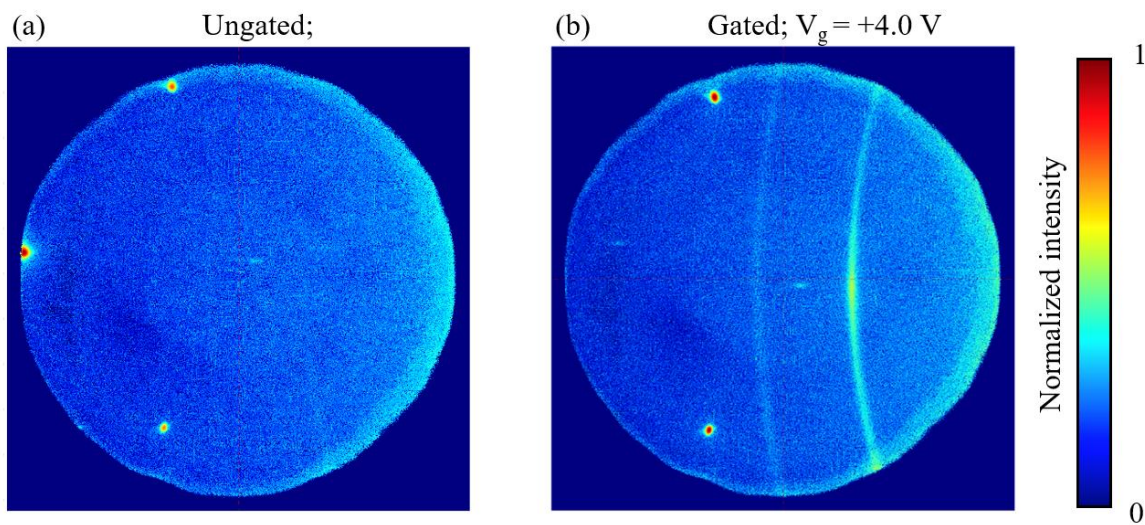

**Fig. S4:** 2D XRD intensity maps obtained from an (a) ungated and a (b) $V_g$ = +4.0 V gated sample.

Figures S4 (a) and (b) display the 2D XRD intensity maps obtained for the ungated and +4.0 V gated samples respectively. Figure S5 displays the corresponding integrated intensities as a function of the scattering angle. Although clear powder rings are observed in the gated sample's intensity map in Fig. S4 (b), these were identified to be that of the Au that was used to form the electrical contacts for transport measurements (see Fig. S5). The Au peaks are observed to have a double-peak shape in the gated sample. No peaks are expected at these scattering angles for any of the sub-sulfide phases. Therefore, the double peak is likely a result of a change in the (apparent) lattice parameters of part of the Au (due to, *e.g.*, strain, contamination, intermixing, height-correction errors, *etc.*). No additional peaks apart from that of the $NiS_2$ phase can be discerned.

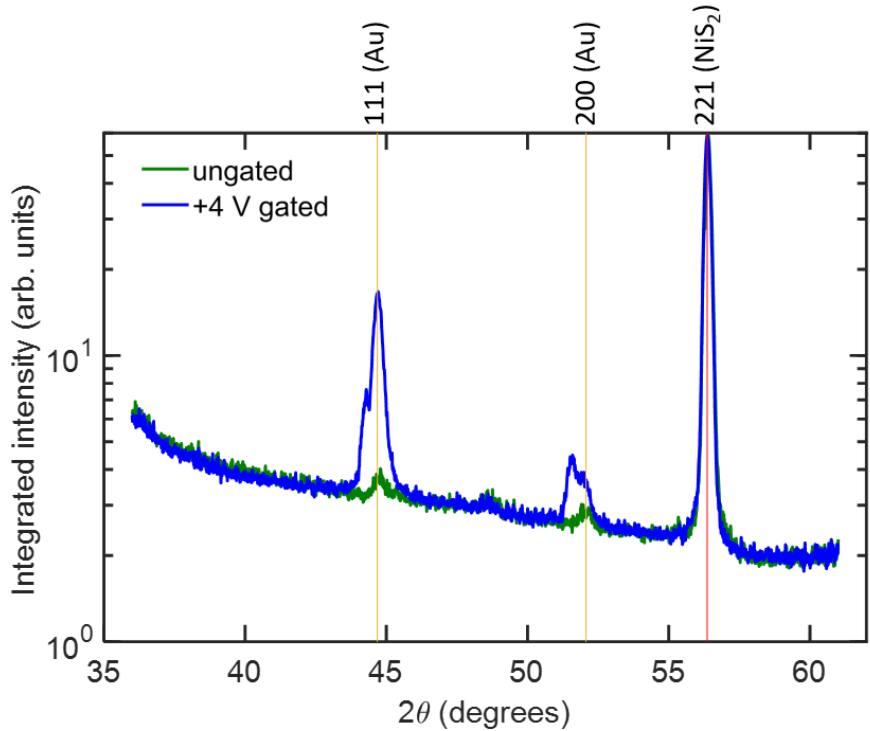

**Fig. S5:** Integrated XRD intensity as a function of the scattering angle $2\theta$ for the same samples as in Fig. S4. The calculated $2\theta$ for Au and $NiS_2$ are indicated as vertical yellow and red lines respectively.

[1] A. A. Ponomareva, V. A. Moshnikov, and G. Suchaneck, IOP Conf. Ser.: Mater. Sci. Eng. **47**, 012052 (2013).